\documentclass[prd,aps,floats,twocolumn]{revtex4}

\usepackage[dvips]{graphicx}
\usepackage{amssymb}
\usepackage{amsmath}
\begin{document}

\title{A new duality relating density perturbations in \\ expanding and
  contracting Friedmann cosmologies}

\author{Latham A. Boyle$^1$,  Paul J. Steinhardt$^{1,2}$
and Neil Turok$^{3}$}

\affiliation{$^1$Department of Physics, Princeton University,
  Princeton, New Jersey 08544, USA \\ $^2$ School of Natural Sciences,
  Institute for Advanced Study, Olden Lane, Princeton, New Jersey
  08540, USA \\
$^3$DAMTP, CMS, Wilberforce Road, Cambridge, CB3\,0WA, UK}

\date{March 2004}

\begin{abstract}
  For a 4--dimensional spatially-flat Friedmann-Robertson-Walker
  universe with a scalar field $\phi(x)$, potential $V(\phi)$ and
  constant equation of state $w=p/\rho$, we show that an expanding
  solution characterized by $\epsilon=3(1+w)/2$ produces the same
  scalar perturbations as a contracting solution with
  $\widehat{\epsilon}=1/\epsilon$.  The same symmetry applies to both
  the dominant and subdominant scalar perturbation modes.  This result
  admits a simple physical interpretation and generalizes to $d$
  spacetime dimensions if we define
  $\epsilon\equiv[(2d-5)+(d-1)w]/(d-2)$.
\end{abstract}
\maketitle

  \section{Introduction}
  \label{intro}

  In inflationary cosmology \cite{inflation}, a nearly scale-invariant
  spectrum of primordial density perturbations is produced as comoving
  scales leave the Hubble horizon during an early burst of accelerated
  expansion \cite{inflation_perts}. In cyclic
  cosmology \cite{cyclic_ekpyrotic}, the \emph{same} perturbation
  spectrum is produced as comoving scales leave the Hubble horizon
  during a period of slow decelerated
  contraction \cite{cyclic_perts,TTS}.  This agreement between two
  physically dissimilar models is unexpected, but not coincidental.
  As we shall show, the relationship between inflation and the cyclic
  model may be viewed as a special case of a surprisingly simple and
  general duality between expanding and contracting cosmologies.
  
  A Friedmann-Robertson-Walker (FRW) universe with a single scalar
  field $\phi$ and potential $V(\phi)$ is a simple yet important
  system.  In particular, it is the canonical 4d effective theory used
  to model the production of density perturbations in both
  inflationary and cyclic cosmology.  Recent results hint at a
  connection between two apparently--unrelated regimes of this model:
  (\emph{i}) expanding, in the $w\rightarrow -1$ limit, and
  (\emph{ii}) contracting, in the $w\rightarrow\infty$ limit.  Here
  $w\equiv p/\rho$ denotes the ratio of pressure to energy density.
  Long-wavelength scale-invariant density perturbations are produced
  in the limits (\emph{i}) and (\emph{ii}) \cite{PHZ_conditions}, and
  the small deviations from scale invariance near these two limits are
  related by the simple substitution $\epsilon\rightarrow 1/\epsilon$,
  where $\epsilon\equiv 3(1+w)/2$ \cite{spectral_tilt}.
  
  The relationship noted in \cite{PHZ_conditions,spectral_tilt}
  between expanding $w\approx -1$ models and contracting $w\gg 1$
  models, turns out to be a special case of a general and exact
  duality relating expanding and contracting models with identical
  perturbation spectra.  In this paper, we derive this duality,
  focusing on the case where $\epsilon$ (or $w$) is time-independent.
  In a companion paper \cite{dual2}, we generalize the discussion to
  the case where $\epsilon$ is time-varying.  When $\epsilon$ is
  constant, or varies sufficiently slowly, the duality is simple: an
  expanding universe characterized by $\epsilon$ produces
  \emph{exactly} the same scalar perturbations as a contracting
  universe characterized by $\widehat{\epsilon}=1/\epsilon$.  This
  duality applies in arbitrary spacetime dimension (not just 3+1); it
  applies for all $w$ (not just the $w\rightarrow -1$ and
  $w\rightarrow\infty$ limits discussed in
  \cite{PHZ_conditions,spectral_tilt}); it applies to all wavelengths
  (not just the long-wavelength limit); it applies to \emph{both} the
  dominant scalar perturbation mode \emph{and} a subdominant remainder
  (which are related to the growing and decaying modes, respectively).
  
  This duality is of general theoretical interest since it provides a
  new relationship between expanding and contracting universes, and
  exposes an unexpected symmetry of cosmological perturbation theory.
  It is also relevant to cosmological models, like the ekpyrotic and
  cyclic \cite{cyclic_ekpyrotic} scenarios, in which perturbations
  produced during a period of contraction are proposed to propagate
  through a bounce into a subsequent expanding phase.
  These models require that the growing-mode long-wavelength
  perturbation spectrum is preserved across the bounce.  This has been
  a controversial matter.  At first, some authors argued that
  growing-mode perturbations produced in a contracting phase must
  match to pure decaying-mode perturbations as one follows them across
  a bounce into an expanding phase \cite{contrascale}.  At heart, this
  conclusion followed from requiring that the bounce corresponds to a
  comoving or constant-energy-density slice.  However, recent five
  dimensional calculations \cite{TTS,newbrand} indicate that comoving
  or constant-energy-density slices are inappropriate for matching,
  since the bounce event (represented in five dimensions by a brane
  collision) is not synchronous in these slices.  Matching on
  collision-synchronous slices results in the propagation of
  growing-mode perturbations across a bounce, with no change in the
  shape of the long-wavelength spectrum.  Other aspects of the cyclic
  and ekpyrotic models have been criticized \cite{kallosh} (see
  \cite{ansc} for replies), and some have argued that a bounce is
  impossible altogether \cite{contrabounce}.
  While the consistency of a bounce remains to be proven, recent
  work has shown that the traditional hazard of chaotic mixmaster
  behavior is strongly suppressed in the contracting phase of the 
  cyclic model \cite{chaos}.  The metric perturbations exhibit 
  ultralocal behavior in which anisotropies remain small, right up 
  to a few Planck times before the bounce.  In this situation, 
  causality suggests that the bounce should not disturb correlations
  on macroscopic scales over which there can be no communication in
  this finite time interval.  Under these conditions, works by 
  several groups \cite{TTS,pro,newbrand} suggest that perturbations
  generated during the contracting phase may pass into the expanding
  phase.  The subject continues to be an area of active research.
  If the latter suggestions are made rigorous, it would give added
  significance to the results presented herein.

  Other dualities have been identified in the literature \cite{Wands,
    Starobinsky, Brustein, scalefacdual, stringdual} that relate
  cosmological solutions and perturbations.  See Section \ref{other}
  for a comparison.  The duality presented here has the distinctive
  property that it relates two solutions that are stable under
  perturbations.  Hence, both solutions can plausibly play a role in
  realistic cosmological models.
  
  The layout of the paper is as follows.  In section \ref{background},
  we introduce the background (unperturbed) model: a spatially-flat
  FRW model with scalar field $\phi$, potential $V(\phi)$, and
  constant $\epsilon$.  In section \ref{scalars}, we briefly review
  scalar perturbation theory, before deriving exact solutions for
  Mukhanov's $u$ and $v$ variables.  We note that $u$ is invariant
  under $\epsilon\rightarrow 1/\epsilon$.  This invariance is
  independent of our vacuum choice for the fluctuations, and provides
  our first glimpse of the duality.  In section \ref{grow_decay} we
  use $u$ and $v$ to separate scalar perturbations into 
pieces that are dominant and subdominant at long wavelengths,
and show that each piece is 
{\it both independently invariant} under $\epsilon \to
  1/\epsilon$.  
We show how the dominant and subdominant pieces relate
  to the scalar perturbation growing and decaying modes.  In section
  \ref{tensors} we consider tensor perturbations.
  In section \ref{other}, we contrast our duality with other kinds of
  cosmological dualities that have been studied in the literature
  \cite{Wands,Starobinsky,Brustein,scalefacdual,stringdual}.
  In section \ref{discuss}, we interpret our duality as a relation
  between the scale factor and the Hubble parameter, discuss its
  observational significance, and mention some open questions.  In an
  appendix, we generalize our results to $d$ spacetime dimensions.

  \section{Background model}
  \label{background}

  A spatially-flat Friedmann-Robertson-Walker (FRW) universe with
  scalar field $\phi$ and potential $V(\phi)$ is described by the
  metric
  \begin{equation}
    \label{frw_metric}
    \textrm{d}s^{2}=a(\tau)^{2}
    \left[-\textrm{d}\tau^{2}+\textrm{d}\vec{x}^{\,2}\right].
  \end{equation}
  The unperturbed scalar field $\phi_{0}(\tau)$ and scale factor
  $a(\tau)$ obey the Friedmann equations
  \begin{subequations}
    \label{friedmann_eqs}
    \begin{eqnarray}
      6(a'^{\,2}/a^{4}) & = & 2\rho\\ 6(a''/a^{3})\; & = & \rho-3p
    \end{eqnarray}
  \end{subequations}
  where we have chosen units such that $c=\hbar=8\pi G=1$, a prime
  ($\,'\,$) denotes a conformal time derivative
  $\textrm{d}/\textrm{d}\tau$, and the energy density and pressure are
  given by
  \begin{subequations}
    \label{rho_p}
    \begin{eqnarray}
      \rho & = & (1/2)a^{-2}\phi_{0}'^{\,2}+V(\phi_{0})
      \\ p & = & (1/2)a^{-2}\phi_{0}'^{\,2}-V(\phi_{0}).
    \end{eqnarray}
  \end{subequations}
  Instead of the usual variable $w\equiv p/\rho$, it will be more
  convenient to use
  \begin{equation}
    \epsilon\equiv 3(1+w)/2.
  \end{equation}
  to parameterize the equation of state.  Equations
  (\ref{friedmann_eqs},\ref{rho_p}) imply $-1\leq w<\infty$ (or
  equivalently $0\leq\epsilon<\infty$).  If $w$ is near $-1$, then
  $\epsilon\ll 1$ is the usual slow-roll parameter; but we make
  \emph{no slow-roll approximation} in this paper, and $\epsilon$ may
  be arbitrarily large.
  
  From now on, we shall assume that $\epsilon$ is constant, not equal
  to unity.  (The self-dual case $\epsilon=1$ possesses special
  behavior which we shall not study here.)  Then the solution of
  equations (\ref{friedmann_eqs},\ref{rho_p}) is:
  \begin{subequations}
    \label{background_soln}
    \begin{eqnarray}
      \label{a(tau)}
      a(\tau) & = & |\tau|^{1/(\epsilon-1)}\\
      \label{phi(tau)}
      \phi_{0}(\tau) & = &
      \pm\frac{(2\epsilon)^{1/2}}{\epsilon-1}\,\textrm{ln}|\tau|\\
      \label{V(phi)}
      V(\phi) & = & \frac{3-\epsilon}{(\epsilon-1)^{2}}\,
      \textrm{exp}\!\!\left[\mp(2\epsilon)^{1/2}\phi\right]
    \end{eqnarray}
  \end{subequations}
  where, to fix integration constants we have, without loss of
  generality, chosen the origin of conformal time so that $a(0)=0$,
  normalized the scale factor so that $a(1)=1$, and redefined
  $\phi_{0}\rightarrow\phi_{0}+$constant so that $\phi_{0}(1)=0$.

  This solution separates into 4 cases:
  \begin{itemize}
  \item (a) expanding, $0\leq\epsilon<1$, $-\infty<\tau<0$;
  \item (b) expanding, $1<\epsilon<\infty$, $0<\tau<\infty$;
  \item (c) contracting, $0\leq\epsilon<1$, $0<\tau<\infty$;
  \item (d) contracting, $1<\epsilon<\infty$, $-\infty<\tau<0$.
  \end{itemize}
  \begin{figure}
    \begin{center}
      \includegraphics[width=3in]{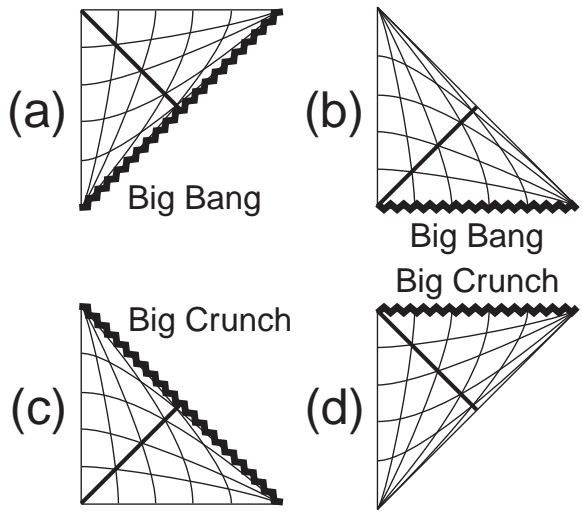}
    \end{center}
    \caption{Penrose diagrams for spatially-flat FRW universes with:
      (a) $0<\epsilon<1$, expanding; (b) $1<\epsilon<\infty$,
      expanding; (c) $0<\epsilon<1$, contracting; (d)
      $1<\epsilon<\infty$, contracting.  The left edge of each diagram
      is the world line of a comoving observer at the origin; curved
      lines represent other comoving world lines and spatial
      hypersurfaces.  The Hubble horizon is a curve connecting the
      $90^{o}$ vertex to the lightlike boundary, but the precise curve
      depends on $\epsilon$.  For illustration, we have shown the
      horizon for $\epsilon=0$ in (a, c) and for $\epsilon=2$ in (b,
      d).  In this paper, we focus on cases (a) and (d), in which
      comoving scales exit the Hubble horizon.}
    \label{penrose}
  \end{figure}
  The Penrose diagrams for these 4 possibilities are shown in Figure
  \ref{penrose}.  Case (b) corresponds to an ordinary expanding FRW
  model with matter or radiation domination, $\epsilon=3/2$ or $2$,
  respectively.  We are interested in the two cases, (a) and (d), in
  which $\tau$ runs from $-\infty\rightarrow 0$ since, in these two
  cases, comoving length scales start \emph{inside} the Hubble horizon
  at early times, and end \emph{outside} the horizon at late times.
  Since we wish to study the amplification of perturbations as modes
  leave the horizon, these are the two relevant cases.  Thus, we will
  always assume $\epsilon<1\Leftrightarrow\textrm{expanding}$ and
  $\epsilon>1\Leftrightarrow\textrm{contracting}$.  The duality
  discussed below pairs solutions of type (a) with solutions of type
  (d).  Figure (\ref{penrose}) emphasizes that (a) and (d) have
  similar causal structure, but different singularities.

  \section{Scalar perturbations}
  \label{scalars}

  In this section, we introduce relevant aspects of gauge-invariant
  scalar perturbation theory in 4 dimensions, and catch our first
  glimpse of the duality discussed in section \ref{grow_decay}. For a
  thorough introduction to gauge-invariant
  perturbations \cite{Bardeen}, see \cite{Kodama_Sasaki} and
  \cite{pert_reviews}.  We work in Fourier space throughout, so every
  perturbation variable carries an implicit subscript $\vec{k}$ which,
  for brevity, is not shown explicitly.  Write the perturbed metric
  \begin{subequations}
    \label{both_perts}
    \begin{eqnarray}
      \label{perturbed_metric}
      \textrm{d}s^{2}/a^{2}&=&-(1+2AY)\textrm{d}\tau^{2}
      -2BY_{i}\textrm{d}\tau\textrm{d}x^{i}\nonumber \\
      &&+\left[(1+2H_{L}Y)\delta_{ij}+2H_{T}Y_{ij}\right]
      \textrm{d}x^{i}\textrm{d}x^{j}
    \end{eqnarray}
    and perturbed scalar field
    \begin{equation}
      \label{perturbed_scalar_field}
      \phi=\phi_{0}(\tau)+\delta\phi(\tau)Y
    \end{equation}
  \end{subequations}
  where $Y(\vec{x})$, $Y_{i}(\vec{x})$, and $Y_{ij}(\vec{x})$ are
  scalar harmonics (see Appendix C in \cite{Kodama_Sasaki}).  The
  corresponding perturbations of the Einstein tensor and
  energy-momentum tensor (see Appendices D and F in
  \cite{Kodama_Sasaki}) are related to one another through the
  perturbed Einstein equations, $\delta G^{\mu}_{\;\;\nu}=\delta
  T^{\mu}_{\;\;\nu}$.

  It is well known that scalar perturbations in a spatially-flat FRW
  universe with scalar field $\phi$ and potential $V(\phi)$ are
  completely characterized by a single gauge-invariant variable.  But
  the choice of this variable is neither unique nor standard; two of
  the most familiar options are the ``Newtonian potential,'' $\Phi$,
  and the ``curvature perturbation,'' $\zeta$.

  The gauge-invariant Newtonian potential $\Phi$ is most easily
  understood in ``Newtonian gauge'' ($B\! =\! H_{T}\! =\! 0$), where
  it is related to the metric perturbations in a simple way:
  $\Phi=A=-H_{L}$.  It obeys the equation of motion
  \begin{equation}
    \label{Phi_EOM}
    \Phi''+2\left[\frac{a'}{a}-\frac{\phi_{0}''}{\phi_{0}'}\right]
    \Phi'+2\left[k^{2}+2\mathcal{H}'-2\mathcal{H}
    \frac{\phi_{0}''}{\phi_{0}'}\right]\Phi=0
  \end{equation}
  where $k=|\vec{k}|$ is the magnitude of the (comoving) Fourier
  3-vector.  On the other hand, the gauge-invariant perturbation
  variable $\zeta$ is most easily understood in ``comoving gauge''
  ($H_{T}=\delta T^{0}_{\;\; i}=0$), where it represents the curvature
  perturbation on spatial-hypersurfaces, and is related to the spatial
  metric perturbation in a simple way: $\zeta=-H_{L}$.  The condition
  $\delta T^{0}_{\;\; i}=0$ also implies that $\delta\phi=0$ in this
  gauge.  $\zeta$ obeys the equation of motion
  \begin{equation}
    \label{zeta_EOM}
    \zeta''+2(z'/z)\zeta'+k^{2}\zeta=0
  \end{equation}
  where $z\equiv a^{2}\phi_{0}'/a'$.  $\Phi$ and $\zeta$ are related
  to each other by
  \begin{subequations}
    \label{zeta_phi_relations}
    \begin{eqnarray}
      \label{zeta_from_phi}
      \zeta & = & \Phi+\frac{1}{\epsilon}
      \left[(a/a')\Phi'+\Phi\right]\\
      \label{Phi_from_zeta}
      \Phi & = & -\epsilon(a'/a)k^{-2}\zeta'\;.
    \end{eqnarray}
  \end{subequations}
  Note that our definitions for $\Phi$ and $\zeta$ agree with those in
  \cite{pert_reviews}.  But beware: in \cite{Kodama_Sasaki}, the
  gauge-invariant Newtonian potential is denoted $\Psi$, while $\Phi$
  denotes a different (though closely related) variable.

  It is convenient to introduce new variables, $u$ and
  $v$ \cite{Mukhanov}, by multiplying $\Phi$ and $\zeta$ by
  $k$-independent functions of $\tau$:
  \begin{equation}
    \label{def_uv}
    u\equiv (a/\phi_{0}')\Phi \qquad v\equiv z\zeta.
  \end{equation}
  Note that $u$ and $v$ have the same $k$-dependence as $\Phi$ and
  $\zeta$, respectively, and may serve as surrogates for $\Phi$ and
  $\zeta$.  If we also define background quantities, $\theta$ and $z$:
  \begin{equation}
    \label{def_theta}
    \theta\equiv 1/z\equiv\frac{a'}{a^{2}\phi_{0}'},
  \end{equation}
  then $u$ and $v$ obey simple equations of motion
  \begin{subequations}
    \label{uv_EOM}
    \begin{eqnarray}
      \label{u_EOM}
      u''+(k^{2}-\theta''/\theta)u & = & 0\\
      \label{v_EOM}
      v''+(k^{2}-z''/z)v & = & 0
    \end{eqnarray}
  \end{subequations}
  and are related to each other by
  \begin{subequations}
    \label{u_v_relations}
    \begin{eqnarray}
      \label{v_from_u}
      kv & = & 2k[u'+(z'/z)u]\\
      \label{u_from_v}
      -ku & = & \frac{1}{2k}
      [v'+(\theta'/\theta)v].
    \end{eqnarray}
  \end{subequations}

  We must choose a vacuum state for the fluctuations, which
  corresponds to specifying appropriate boundary conditions for $u$
  and $v$ (see Ch.3 in Birrel\&Davies \cite{Birrell_Davies}).  The
  standard choice is the Minkowski vacuum of a comoving observer in
  the far past (when all comoving scales were far inside the Hubble
  horizon), corresponding to the boundary conditions
  \begin{subequations}
    \label{vacuum_def}
    \begin{eqnarray}
      \label{u_bc}
      u & \rightarrow i(2k)^{-3/2}\textrm{e}^{-ik\tau}\\
      \label{v_bc}
      v & \rightarrow \;(2k)^{-1/2}\textrm{e}^{-ik\tau}
    \end{eqnarray}
  \end{subequations}
  as $\tau\rightarrow-\infty$.  Using (\ref{u_v_relations}), it is
  easy to check that these two boundary conditions are equivalent.

  When $\epsilon$ is time-independent, we can use
  (\ref{background_soln}) to find
  \begin{subequations}
    \begin{eqnarray}
      \label{theta_eps}
      \theta''/\theta & = &
      \frac{\epsilon}{(\epsilon-1)^{2}\tau^{2}}\\
      \label{z_eps}
      z''/z & = &
      \frac{2-\epsilon}{(\epsilon-1)^{2}\tau^{2}}
    \end{eqnarray}
  \end{subequations}
  Then we may solve (\ref{uv_EOM}) to obtain
  \begin{subequations}
    \begin{eqnarray}
      u(x)=x^{1/2}\left[A^{(1)}H^{(1)}_{\alpha}(x)
    +A^{(2)}H^{(2)}_{\alpha}(x)\right]\\
      v(x)=x^{1/2}\left[B^{(1)}H^{(1)}_{\beta}(x)
    +B^{(2)}H^{(2)}_{\beta}(x)\right]
    \end{eqnarray}
  \end{subequations}
  where $x\equiv k|\tau|$ is a dimensionless time variable,
  $A^{(1,2)}$ and $B^{(1,2)}$ are constants, $H^{(1,2)}_{\rho}(x)$ are
  Hankel functions, and we have defined
  \begin{subequations}
    \begin{eqnarray}
      \label{def_alpha}
      \alpha \equiv & \sqrt{(\theta''/\theta)\tau^{2}+1/4} &
      =\frac{1}{2}\left|\frac{\epsilon+1}{\epsilon-1}\right|\\
      \label{def_beta}
      \beta \equiv & \sqrt{(z''/z)\tau^{2}+1/4} &
      =\frac{1}{2}\left|\frac{\epsilon-3}{\epsilon-1}\right|
    \end{eqnarray}
  \end{subequations}
  In the far past ($x\rightarrow\infty$) we use the asymptotic Hankel
  expression,
  \begin{equation}
    \label{hankel_asymptotic}
      H^{(1,2)}_{s}(x)\rightarrow\sqrt{\frac{2}{\pi x}}
    \textrm{exp}\left[\pm\,
    i\left(x-\frac{s\,\pi}{2}-\frac{\pi}{4}\right)\right]
  \end{equation}
  so the boundary conditions (\ref{vacuum_def}) imply
  \begin{subequations}
    \begin{eqnarray}
      \label{u_soln}
      u & = & \frac{\mathcal{P}_1}{2k}(\pi x/4k)^{1/2}
      H_{\alpha}^{(1)}(x)\\
      \label{v_soln}
      v & = & \,\mathcal{P}_{2}\,(\pi x/4k)^{1/2}
      H_{\beta}^{(1)}(x)
    \end{eqnarray}
  \end{subequations}
  where
  \begin{subequations}
    \begin{eqnarray}
      \label{P_1}
      \mathcal{P}_1=\textrm{exp}[i(2\alpha+3)\pi/4]\\
      \label{P_2}
      \mathcal{P}_{2}=\textrm{exp}[i(2\beta+1)\pi/4]
    \end{eqnarray}
  \end{subequations}
  are $k$-independent complex phase factors.

  Note from (\ref{theta_eps}) that the equation of motion for $u$,
  (\ref{u_EOM}), is invariant under $\epsilon\rightarrow 1/\epsilon$,
  while the boundary condition, (\ref{u_bc}), is independent of
  $\epsilon$.  As a result, our expressions (\ref{def_alpha}) for
  $\alpha$ and (\ref{u_soln}) for $u$ are invariant under $\epsilon
  \rightarrow 1/\epsilon$.  This is our first glimpse of the duality
  discussed below.

  We stress that this result does not depend on the particular vacuum
  choice (\ref{u_bc}).  Any boundary condition that is independent of
  $\epsilon$ (or, more generally, invariant under $\epsilon\rightarrow
  1/\epsilon$) will work.  And it is natural to expect the boundary
  condition to be independent of $\epsilon$, since it is imposed in
  the far past, when comoving scales are far inside the Hubble
  horizon.
  
  \section{Dominant and subdominant modes}
  \label{grow_decay}

  In this section, we show that $u$ and $v$ can be decomposed into
  pieces that are dominant and subdominant at long wavelengths such
  that each is invariant under the transformation $\epsilon\rightarrow
  1/\epsilon$.  The dominant and subdominant parts are closely related
  growing and decaying modes over the range of $w$ relevant to
  cosmological model-building, as we explain below.
  
  For this analysis, it is convenient to scale $u$ and $v$ by
  appropriate powers of $|\tau|$, so that they only depend on $k$ and
  $\tau$ through the dimensionless combination $x=k|\tau|$.  Thus,
  using (\ref{u_soln}, \ref{v_soln}), define
  \begin{subequations}
    \label{def_uv_bar}
    \begin{eqnarray}
      \label{def_u_bar}
      \bar{u} \equiv |\tau|^{-3/2}u & = &
      \frac{\mathcal{P}_1}{2x}(\pi/4)^{1/2}H_{\alpha}^{(1)}(x)\\
      \label{def_v_bar}
      \bar{v} \equiv |\tau|^{-1/2}v & = &
      \,\mathcal{P}_{2}\,(\pi/4)^{1/2}H_{\beta}^{(1)}(x).
    \end{eqnarray}
  \end{subequations}
  Note that $\bar{u}$ and $\bar{v}$ have the same $k$-dependence as
  $u$ and $v$, respectively, or $\Phi$ and $\zeta$, respectively, and
  may be used in place of these more standard variables.  

  
  To make the meaning of ``dominant'' and ``subdominant'' precise,
  consider two linearly independent functions $f_{1}(x)$ and
  $f_{2}(x)$.  If $\lim_{x \to 0} f_{2}(x)/f_{1}(x)$ exists, then
  $f_{1}(x)$ and $f_{2}(x)$ can be related by a linear transformation
  to two new functions, $f_{{\rm dom}}(x)$ and $f_{{\rm sub}}(x)$,
  satisfying
  \begin{equation}
    \label{fsub_over_fdom}
    \lim_{x \to 0} f_{{\rm sub}}(x)/f_{{\rm dom}}(x) = 0.
  \end{equation}
  So the subdominant piece, $f_{{\rm sub}}(x)$, becomes negligible
  relative to the dominant piece, $f_{{\rm dom}}(x)$, for small $x$
  ({\it i.e.} far outside the horizon).  The condition
  (\ref{fsub_over_fdom}) uniquely determines the subdominant piece (up to
  an overall normalization constant) to be
  \begin{equation}
    \label{def_fsub}
    f_{{\rm sub}}(x) = f_{2}(x)-\left[\lim_{y \to 0} 
      \frac{f_{2}(y)}{f_{1}(y)}\right]f_{1}(x),
  \end{equation}
  but does not uniquely fix the dominant piece.  Rather, $f_{{\rm
      dom}}(x)$ may be any linear combination of $f_{1}(x)$ and
  $f_{2}(x)$ that is linearly independent of $f_{{\rm sub}}(x)$.  We
  can now choose $f_{1}(x)=\bar{u}(x)$ and $f_{2}(x)=\bar{v}(x)$, and
  find the corresponding dominant and subdominant functions.
  
  Let us first calculate $f_{{\rm sub}}(x)$.  To compute $\lim_{x \to
    0} f_{2}(x)/f_{1}(x)$, recall the Hankel identity
  \begin{equation}
    \label{hankel_small_x}
    H_{s}^{(1)}(x)\rightarrow -i[\Gamma(s)/\pi]
    (x/2)^{-s}
    \quad\textrm{as}\quad x\rightarrow 0,
  \end{equation}
  where $s>0$ and $\Gamma(s)$ is the Euler gamma function.  Then from
  (\ref{def_u_bar},\ref{def_v_bar}) we find
  \begin{equation}
    \label{f2_over_f1}
    \lim_{x \to 0} \frac{f_{2}(x)}{f_{1}(x)}=\left\{\begin{array}{ll}
        (\mathcal{P}_{2}/\mathcal{P}_{1})4\alpha\quad & 0\leq\epsilon<1\\
        0\quad & 1<\epsilon<\infty
      \end{array}\right.
  \end{equation}
  where we have used the fact that $\alpha+1=\beta$ when $\epsilon<1$,
  while $\alpha+1>\beta$ when $\epsilon>1$.  Now, substituting
  (\ref{def_u_bar}), (\ref{def_v_bar}) and (\ref{f2_over_f1}) into
  (\ref{def_fsub}), using the Hankel identities
  \begin{subequations}
    \begin{eqnarray}
      & H_{s-1}^{(1)}(x)+H_{s+1}^{(1)}(x)
      =\frac{2 s}{x}H_{s}^{(1)}(x)\\
      & H_{-s}^{(1)}(x)
      =\textrm{e}^{i\pi s}H_{s}^{(1)}(x),
    \end{eqnarray}
  \end{subequations}
  and paying careful attention to absolute value signs, we find
  \begin{equation}
    \label{f_sub}
    f_{{\rm sub}}(x)=\mathcal{P}_{3}(\pi/4)^{1/2}H_{\gamma}^{(1)}(x),
  \end{equation}
  where we have defined
  \begin{equation}
    \label{def_gamma}
    \gamma\equiv |\alpha-1|, 
  \end{equation}
  and $\mathcal{P}_{3}=\textrm{exp}[i(2\gamma+1)\pi/4]$ is a
  $k$-independent complex phase factor.  
  
  Now let us turn to $f_{{\rm dom}}(x)$.  The fact that $\lim_{x \to
    0} f_{{\rm sub}}(x)/ \bar{v}(x)=0$ when $\epsilon<1$ shows that
  the dominant mode contributes to $\bar{v}$ in an expanding universe.
  Since  $f_{{\rm sub}}(x) =\bar{v}(x)$ when $\epsilon>1$, 
 $\bar{v}$ is purely subdominant ({\it i.e.} contains no
  dominant-mode contribution) in a contracting universe.  By contrast,
  $\bar{u}$ and $f_{s}(x)$ are always linearly independent, and hence
  $\lim_{x\to 0}f_{{\rm sub}} (x)/\bar{u}(x)=0$ for all $\epsilon$.  
  Thus, we can use the freedom in defining $f_{{\rm dom}}$ to choose
  \begin{equation}
    \label{f_dom}
    f_{{\rm dom}}(x)=\bar{u}(x)=
    \frac{\mathcal{P}_1}{2x}(\pi/4)^{1/2}H_{\alpha}^{(1)}(x)
  \end{equation}
  for the dominant piece.
  
  Notice that the expressions (\ref{f_sub}) for $f_{{\rm sub}}(x)$ and
  (\ref{f_dom}) for $f_{{\rm dom}}(x)$ are {\it both} invariant under
  $\epsilon\rightarrow 1/\epsilon$, because $\alpha$ is invariant.
  Thus, we see that the subdominant mode {\it automatically} possesses
  this symmetry, since $f_{{\rm sub}}(x)$ is uniquely determined (up
  to a normalization factor) by the condition (\ref{fsub_over_fdom}).
  Furthermore, we have shown that we can choose a linear combination
  of $\bar{u}(x)$ and $\bar{v}(x)$ such that $f_{{\rm dom}}(x)$
  displays the same symmetry.  
(If we had made the wrong
  choice for $f_{{\rm dom}}(x)$, then the exact $\epsilon\rightarrow
  1/\epsilon$ symmetry of the dominant mode would be hidden, and would
  only re-appear in the long-wavelength limit.)
  
\begin{figure}
    \begin{center}
      \includegraphics[width=3in]{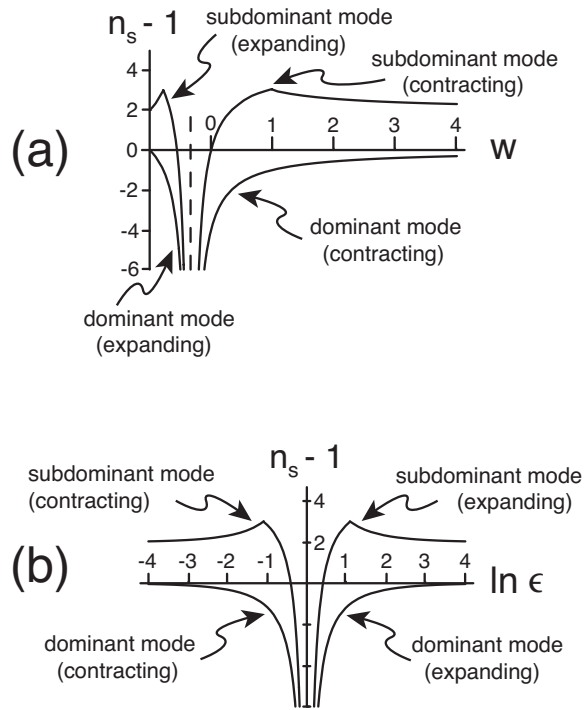}
    \end{center}
    \caption{The dominant and subdominant scalar spectral
      indices (a) as a function of $w$ and (b) as a function of
      $\textrm{ln}\epsilon$.  Note especially the symmetry of (b).}
    \label{bothspec}
  \end{figure}

  In the $x\rightarrow 0$ limit, define dominant and subdominant
  scalar spectral indices, $n_{{\rm dom}}$ and
  $n_{{\rm sub}}$, which satisfy
  \begin{subequations}
    \label{def_n_dom_n_sub}
    \begin{eqnarray}
      x^{3}|f_{{\rm dom}}|^{2}   & \propto & x^{n_{{\rm dom}}-1}\\
      x^{3}|\,f_{{\rm sub}}\,|^{2} & \propto & x^{n_{{\rm sub}}-1}
    \end{eqnarray}
  \end{subequations}
  Using (\ref{f_sub}) and (\ref{f_dom}), along with the Hankel
  identity (\ref{hankel_small_x}), we find
  \begin{subequations}
    \begin{eqnarray}
      \label{n_dom}
      n_{{\rm dom}}-1\!\! & = 1-2\alpha =\! 
      & 1-\;\Big|\frac{\epsilon+1}{\epsilon-1}\Big|\\
      \label{n_sub}
      n_{{\rm sub\;}}-1\!\! & = 3-2\gamma =\!
      & 3-\left|\Big|\frac{\epsilon+1}{\epsilon-1}\Big|-2\right|
    \end{eqnarray}
  \end{subequations}
  These spectral indices are plotted in Fig. \ref{bothspec}a (as a
  function of $w$), and in Fig. \ref{bothspec}b (as a function of
  $\epsilon$).  Again, notice that $n_{{\rm dom}}$ and $n_{{\rm sub}}$
  are \emph{both} invariant under $\epsilon\rightarrow 1/\epsilon$.
  This symmetry is manifest in Fig. \ref{bothspec}b.
  
  Since $\epsilon$ lies in the range $0\leq\epsilon<\infty$, this
  duality formally pairs every expanding ($\epsilon<1$) universe with
  a contracting ($\widehat{\epsilon}>1$) universe, and \emph{vice
    versa}.  However, the background solution (\ref{background_soln})
  is only stable against small perturbations in two cases: (\emph{i})
  expanding with $\epsilon<1$ or (\emph{ii}) contracting with
  $\epsilon>3$ \cite{PHZ_conditions,chaos}.  Thus, an expanding model
  and its contracting dual are \emph{both} stable when $\epsilon<1/3$
  ($w<-7/9$) and $\widehat{\epsilon}>3$ ($\widehat{w}>1$).  Also note,
  in agreement with \cite{PHZ_conditions}, that there are only two
  limits in which an approximately scale-invariant
  ($n_{\textrm{grow}}-1\approx 0$) spectrum of scalar perturbations is
  produced: (\emph{i}) when $\epsilon\rightarrow 0$ ($w\rightarrow
  -1$), corresponding to the inflationary regime and (\emph{ii}) when
  $\epsilon\rightarrow\infty$ ($w\rightarrow\infty$) corresponding to
  the cyclic/ekpyrotic regime.
  
  The dominant and subdominant pieces, $f_{{\rm dom}}(x)$ and $f_{{\rm
      sub}}(x)$ are related to the growing and decaying modes of
  $\bar{u}$ and $\bar{v}$, which may simply be obtained by replacing
  the Hankel functions in (\ref{def_u_bar},\ref{def_v_bar}) with the
  corresponding Neumann functions or Bessel functions:
  \begin{subequations}
    \label{uv_grow_decay}
    \begin{eqnarray}
      \bar{u}_{{\rm grow}}\propto x^{-1}Y_{\alpha}(x) & &
      \bar{u}_{{\rm decay}}\propto x^{-1}J_{\alpha}(x) \\
      \bar{v}_{{\rm grow}}\propto Y_{\beta}(x) \!\!\qquad & & 
      \bar{v}_{{\rm decay}}\propto J_{\beta}(x). 
    \end{eqnarray}
  \end{subequations}
  We may define growing-mode and decaying-mode spectral indices for
  $\bar{u}$ and $\bar{v}$, just as we did for $f_{{\rm sub}}$ and
  $f_{{\rm dom}}$ in (\ref{def_n_dom_n_sub}).  Now restrict attention
  to the ranges $\epsilon<1/3$ and $\widehat{\epsilon}>3$ ({\it i.e.}
  the range over which the duality pairs stable expanding models to
  stable contracting models).  Then, using (\ref{uv_grow_decay}) along
  with the (\ref{hankel_small_x}), we find
  \begin{itemize}
  \item The growing-mode and decaying-mode spectral indices associated
    with $\bar{u}$ are simply equal to $n_{{\rm dom}}$ and $n_{{\rm
        sub}}$, respectively.
  \item The growing-mode spectral index associated with $\bar{v}$ is
    equal to $n_{{\rm dom}}$ in an expanding ($\epsilon<1/3$) universe, 
    but is equal to $n_{{\rm sub}}$ in a contracting
    ($\widehat{\epsilon}>3$) universe.
  \end{itemize}  
  

  \section{Tensor perturbations}
  \label{tensors}
  
  Tensor perturbations are much simpler than scalar perturbations.
  The perturbed metric is
  \begin{equation}
    \label{tensor_perturbed_metric}
    \textrm{d}s^{2}/a^{2}=-\textrm{d}\tau^{2}
    +[\delta_{ij}+2h_{T}Y^{(2)}_{ij}]
    \textrm{d}x^{i}\textrm{d}x^{j}
  \end{equation}
  where $Y^{(2)}_{ij}$ is a tensor harmonic (see Appendix C in
  \cite{Kodama_Sasaki}).  The tensor perturbation $h_{T}$ is
  gauge-invariant and obeys
  \begin{equation}
    \label{tensor_EOM}
    h_{T}''+2(a'/a)h_{T}'+k^{2}h_{T}=0.
  \end{equation}
  It is useful to define a new variable $f_{T}\equiv a h_{T}$ which
  obeys
  \begin{equation}
    \label{f_EOM}
    f_{T}''+(k^{2}-a''/a)f_{T}=0.
  \end{equation}
  Again, the standard vacuum choice is the Minkowski vacuum of a
  comoving observer in the far past, corresponding to the boundary
  condition
  \begin{equation}
    \label{tensor_bc}
    f_{T}\rightarrow (2k)^{-1/2}e^{-ik\tau}
    \qquad\textrm{as}\quad\tau\rightarrow -\infty.
  \end{equation}
  We can now solve for $f_{T}$, just as in the scalar case.  But it is
  simpler to notice that (\ref{background_soln}) and (\ref{def_theta})
  imply $z(\tau)\propto a(\tau)$, and hence $z''/z=a''/a$ when
  $\epsilon$ is constant.  Thus, since $v$ and $f_{T}$ obey identical
  equations of motion (compare (\ref{v_EOM}) with (\ref{f_EOM})) and
  boundary conditions (compare (\ref{v_bc}) with (\ref{tensor_bc})),
  we find
  \begin{equation}
    \label{f_soln}
    f_{T}=v=\mathcal{P}_{2}(\pi x/4k)^{1/2}H_{\beta}^{(1)}(x).
  \end{equation}

  The tensor spectral index is defined in the long-wavelength limit by
  $k^{3}|f_{T}|^{2}\propto k^{n_{T}}$.  Using (\ref{hankel_small_x})
  and (\ref{f_soln}) we find
  \begin{equation}
  n_{T}=3-2\beta=3-\left|\frac{\epsilon-3}{\epsilon-1}\right|.
  \end{equation}

  Note, in particular, that this expression is \emph{not} invariant
  under $\epsilon\rightarrow 1/\epsilon$.  An expanding universe with
  equation of state $\epsilon$ produces a tensor spectrum which is
  much redder than the tensor spectrum produced in a contracting
  universe with $\widehat{\epsilon}=1/\epsilon$: $n_{T}\leq
  \widehat{n}_{T}-2$.  

  \section{Other dualities}
  \label{other}

  It is interesting to contrast our duality with other cosmological
  dualities that have been discussed in the literature.  
  
  One duality, due to Wands \cite{Wands} (see also
  \cite{Starobinsky}), pairs models that share the same ``$v$''
  perturbations.  By contrast, our duality pairs models that share the
  same ``$u$'' perturbations.  (Note: the variable called ``$u$'' in
  \cite{Wands} is called ``$v$'' in our paper, in agreement with
  Mukhanov's convention \cite{pert_reviews,Mukhanov}.) 
  Thus, whereas our duality connects expanding and contracting models
  through the substitution $\epsilon\rightarrow\widehat{\epsilon} =
  1/\epsilon$ (which leaves $\theta''/\theta$, and hence $u$,
  invariant), Wands' duality instead uses the substitution
  $\epsilon\rightarrow \widehat{\epsilon} = (2\epsilon-3) /
  (\epsilon-2)$ (which leaves $z''/z$, and hence $v$, invariant).  
  For example, his duality pairs an expanding inflationary solution
  ($\epsilon=0$, $w=-1$) with a contracting dustlike solution
  ($\widehat{\epsilon}=3/2$, $\widehat{w}=0$).  Another difference
  between our duality and Wands' stems from the fact that $v$ is
  purely subdominant ({\it i.e.} contains no dominant-mode
  contribution) in a contracting universe (see section
  \ref{grow_decay}).
  Thus, whereas our duality maps the expanding-phase dominant mode to
  the contracting-phase dominant mode, and the expanding-phase
  subdominant mode to the contracting-phase subdominant mode, Wands'
  duality instead associates the expanding-phase dominant mode with
  the contracting-phase subdominant mode.
  
  A second interesting duality, discussed by Brustein \emph{et al}.\
  \cite{Brustein}, applies to a broad class of cosmological
  perturbations.  Associated with each type of perturbation is a
  ``pump''---a particular function of the background fields.  The
  Hamiltonian governing a given perturbation is invariant under a
  transformation that swaps the perturbation with its conjugate
  momentum, and simultaneously inverts the appropriate pump
  function \cite{Brustein}. It is instructive to apply this duality to
  the $u$ and $v$ variables considered in this paper.  In this case,
  the Brustein {\it et al}.\
  duality associates an expanding solution characterized by $\theta$
  (or $\epsilon$) to a contracting universe characterized by
  $\widehat{\theta} = 1/\theta=z$ (or $\widehat{\epsilon} =
  2-\epsilon$).  The transformation $\epsilon\rightarrow
  \widehat{\epsilon}=2-\epsilon$ effectively swaps the variables $u$
  and $v$
  \begin{subequations}
    \begin{eqnarray}
      u\rightarrow \widehat{u}=(i/2k)v \\
      v\rightarrow \widehat{v}=(2k/i)u
    \end{eqnarray}
  \end{subequations}
  as may be verified from (\ref{u_soln},\ref{v_soln}).
  
  Recall that $\epsilon$ lies in the range $0\leq\epsilon<\infty$.
  Thus, our duality formally pairs every expanding ($0\leq\epsilon<1$)
  solution with a contracting ($1<\widehat{\epsilon}<\infty$)
  solution, and \emph{vice versa}.  By contrast, Wands' duality
  relates every expanding solution to a contracting solution with $1 <
  \widehat{\epsilon} \leq 3/2$; but contracting solutions with
  $\widehat{\epsilon}>3/2$ have no expanding dual.  Similarly,
  Brustein \emph{et al.}'s duality pairs every expanding solution with
  a contracting solution in the range $1<\widehat{\epsilon}\leq 2$;
  but contracting solutions with $\widehat{\epsilon}>2$ have no
  expanding dual.
  
  The constant-$\epsilon$ background solutions (\ref{background_soln})
  are only practically useful if they are dynamically stable.  Recall
  that the contracting solutions are stable if and only if
  $\epsilon>3$ ($w>1$) \cite{PHZ_conditions,chaos}. Thus, Wands' and
  Brustein \emph{et al.}'s dualities relate every expanding solution
  to an \emph{unstable} contracting solution.  By contrast, our
  duality relates \emph{stable} expanding solutions with
  $0\leq\epsilon<1/3$ ($-1\leq w<-7/9$) to \emph{stable} contracting
  solutions with $3<\widehat{\epsilon}<\infty$
  ($1<\widehat{w}<\infty$).  In terms of the spectral index, this
  means that $n_{s}>0$ may be produced \emph{either} by a stable
  expanding phase \emph{or} by a stable contracting phase.  In the
  real universe, the condition $n_{s}>0$ is satisfied (experiments
  favor $n_{s}\approx 1$), so that our duality is of practical
  relevance in cosmological model building.
  Some properties of the three different duality relations are
  summarized in Table I.
  
  \begin{table}
    \label{summary}
    \begin{tabular}{|l|l|l|c|c|}
      \hline
      \multicolumn{2}{|c|}{Transformation of} & 
      \multicolumn{1}{|c|}{Range} & Maps stable & \\
      background & perturbations & 
      \multicolumn{1}{|c|}{of $\epsilon$} & to stable?  & Ref. \\
      \hline \hline
      $\epsilon\rightarrow 1/\epsilon\;\;\,$ & $u\rightarrow u$ & 
      $[0,\infty)$ & Yes, if $n_{s}>0$ & - \\
      \hline  \hline
      $\epsilon\rightarrow \frac{2\epsilon-3}{\epsilon-2}\,$ &
      $v\rightarrow v$ & $[0,3/2]$ & never & \cite{Wands,Starobinsky} \\
      \hline
      $\epsilon\rightarrow 2-\epsilon$ & 
      $u\leftrightarrow(i/2k)v$ & $[0,2]$ & never & \cite{Brustein} \\
      \hline
    \end{tabular}
    \caption{Comparison of the duality presented here with
      those presented by Wands \cite{Wands,Starobinsky}
      and by Brustein {\it et al}.\ \cite{Brustein}. The first 
      two columns show how the background and perturbation
      variables transform under each duality.  
      The third column shows the range of $\epsilon$ to which the
      duality applies.
      The fourth column indicates the condition under which 
      an expanding background solution and its
      contracting dual are both stable under small perturbations.}
  \end{table}
  
  Finally, a number of authors have discussed cosmological symmetries
  of the low-energy string effective action.  If one neglects all
  fields in this action besides the dilaton and the metric, then there
  is a well--known ``scale--factor duality'' \cite{scalefacdual}:
  starting with any cosmological solution, one can use this duality to
  generate new solutions.  If, in addition to the dilaton, one
  includes other fields (axions, moduli,\ldots), then the cosmological
  solutions may display more general dualities, and the resulting
  perturbation spectra may be invariant under these dualities
  \cite{stringdual}. But note that these symmetries typically relate
  different solutions of a single effective action.  By contrast,
  the dualities in 
  Table I relate two different cosmological background 
  solutions corresponding to \emph{two different Lagrangians}: an
  expanding universe, with potential $V(\phi)$, is dual to a
  contracting universe, with a \emph{different} potential
  $\widehat{V}(\phi)$.

  \section{Discussion}
  \label{discuss}  

  Beyond its inherent theoretical interest, our duality may be
  observationally relevant if (as discussed in section \ref{intro})
  long-wavelength correlations produced during a contracting phase
  successfully propagate into a subsequent expanding phase.  This
  suggests a fundamental degeneracy: an ideal measurement of the
  ``primordial" scalar perturbation spectrum may be unable to
  determine whether the perturbations were generated by an expanding
  phase or by its contracting dual.  Luckily, as shown in section
  \ref{tensors}, tensor perturbations break this degeneracy: an
  expanding model produces a much redder tensor spectrum than its
  contracting dual.  In particular, a detection of tensors in the
  cosmic microwave background would indicate that these perturbations
  were generated in an expanding phase, since the dual contracting
  phase would produce an undetectably small tensor spectrum on these
  cosmological length scales \cite{gravity_waves}.

  Using the background solutions (\ref{background_soln}), we may think
  of the duality as relating two different scale factors, $a(\tau)$
  and $\widehat{a}(\tau)$, or two different scalar potentials,
  $V(\phi)$ and $\widehat{V}(\phi)$.  Alternatively, recall that
  $a\propto |t|^{1/\epsilon}$ and $H^{-1}\equiv a^{2}/a'\propto t$,
  where $t$ is the proper time of a comoving observer and $H^{-1}$ is
  the ``Hubble scale."  Thus, two dual universes are related by
  \begin{equation}
    \label{dln/dln_correspondence}
    \textrm{d\,ln}H/\textrm{d\,ln}a
    =\textrm{d\,ln}\widehat{a}/\textrm{d\,ln}\widehat{H}.
  \end{equation}
  
  So the duality effectively swaps the scale factor with the Hubble
  scale, and simultaneously swaps expansion and contraction.  For
  example, in the $\epsilon\rightarrow 0$ ($w\rightarrow -1$) limit,
  the scale factor grows rapidly while the Hubble length grows slowly;
  whereas in the $\epsilon\rightarrow\infty$ ($w\rightarrow\infty$)
  limit, the Hubble length shrinks rapidly while the scale factor
  shrinks slowly.  Expanding models in which modes exit the horizon
  most rapidly ($w\rightarrow -1$) and most slowly ($w\rightarrow
  -1/3$), are associated with contracting models in which modes exit
  most rapidly ($w\rightarrow\infty$) and most slowly ($w\rightarrow
  -1/3$), respectively.
  
  Finally, let us mention several open questions.  We have studied the
  duality using a simple model---spatially-flat FRW spacetime with a
  single scalar field $\phi$; what happens in more complicated models?
  We have used linear perturbation theory; what happens in the
  nonlinear regime?  We have restricted ourselves to time-independent
  $\epsilon$; what if we allow time-varying $\epsilon$?  This final
  question is treated in a companion paper \cite{dual2}.

  \begin{acknowledgments}
    LB thanks A. Starobinsky for pointing out helpful references.  LB
    is supported by an NSF Graduate Research Fellowship.  PJS is
    supported in part by US Department of Energy Grant
    DE-FG02-91ER40671. PJS is also Keck Distinguished Visiting
    Professor at the Institute for Advanced Study with support from
    the Wm.~Keck Foundation and the Monell Foundation.  The work of NT
    is supported by PPARC (UK).
  \end{acknowledgments}

  \appendix*

  \section{Generalization to arbitrary spacetime dimension}
  \label{d_dimensions}

  In sections \ref{background}---\ref{tensors}, we restricted our
  discussion to 4 spacetime dimensions.  Here we sketch the
  generalization to $d$ spacetime dimensions, for $d\geq 4$.  To
  understand this appendix, read sections
  \ref{background}---\ref{tensors} first.  Gauge-invariant
  perturbation theory in $d$ spacetime dimensions is treated
  thoroughly in \cite{Kodama_Sasaki} (see especially the appendices
  therein).

  The background metric (\ref{frw_metric}) obeys the Friedmann
  equations
  \begin{subequations}
    \label{new_friedmann_eqs}
    \begin{eqnarray}
    (d-1)(d-2)a'^{\,2}\!/a^4=2\rho
      \qquad\qquad\qquad\qquad\;\\
    (d-1)(d-2)\,a''/a^3=(5-d)\rho-(d-1)p\quad
    \end{eqnarray}
  \end{subequations}
  where $\rho$ and $p$ are given by (\ref{rho_p}).  Instead of
  $w\equiv p/\rho$, parameterize the equation of state with
  \begin{equation}
    \label{new_def_eps}
    \epsilon\equiv\frac{(2d-5)+(d-1)w}{d-2}.
  \end{equation}
  From eqs. (\ref{rho_p}, \ref{new_friedmann_eqs}) we find $-1\leq
  w<\infty$ or $\frac{d-4}{d-2}\leq\epsilon<\infty$.  For constant
  $\epsilon$, the solution of (\ref{rho_p}, \ref{new_friedmann_eqs})
  is
  \begin{subequations}
    \label{new_background_soln}
    \begin{eqnarray}
      \label{new_a(tau)}
      && \!\!\!\!\!\!\!\!\!\!\!\!\!\!\!\!\!\!\!\! a(\tau)=
      |\tau|^{2/[(d-2)(\epsilon-1)]}\\
      \label{new_phi(tau)}
      && \!\!\!\!\!\!\!\!\!\!\!\!\!\!\!\!\!\!\!\!\phi_{0}(\tau)=
      \pm\frac{1}{\epsilon-1}
      \sqrt{2\left(\epsilon-\frac{d-4}{d-2}\right)}\;
      \textrm{ln}|\tau|\\
      \label{new_V(phi)}
      && \!\!\!\!\!\!\!\!\!\!\!\!\!\!\!\!\!\!\!\!V(\phi)=
      \frac{3-\epsilon}{(\epsilon-1)^{2}}\,
      \textrm{exp}\!\!\left[\mp\sqrt{2\left(\epsilon-
      \frac{d-4}{d-2}\right)}\;\phi\right]
    \end{eqnarray}
  \end{subequations}
  where we have chosen $a(0)=0$, $a(1)=1$ and $\phi_{0}(1)=0$.

  As shown in \cite{Kodama_Sasaki}, perturbations in $d$ spacetime
  dimensions may be decomposed into scalars, vectors and tensors, just
  as in 4 dimensions, and gauge-invariant variables may be defined.
  In particular, we can again introduce scalar perturbations through
  equations (\ref{perturbed_metric}, \ref{perturbed_scalar_field}),
  and describe these perturbations with a single gauge-invariant
  variable.  The gauge-invariant Newtonian potential, $\Phi$, is most
  easily understood in ``Newtonian gauge'' ($B=H_{L}=0$), where it is
  related to the metric perturbations in a simple way:
  $\Phi=A=-(d-3)H_{L}$.  The gauge-invariant variable $\zeta$ is most
  easily understood in ``comoving gauge'' ($H_{T}=\delta
  T^{0}_{i}=0$), where it is related to the spatial metric
  perturbation in a simple way: $\zeta=-H_{L}$.  Note also that
  $\delta\phi=0$ in this gauge.  If we take the $d$-dimensional
  definitions of $u$, $v$, $\theta$ and $z$ to be:
  \begin{equation}
    \label{new_def_uv}
    u\equiv\frac{1}{2}\left(\frac{d-2}{d-3}\right)
    \frac{a^{(d-2)/2}}{\phi_{0}'}\Phi
    \qquad\qquad v\equiv z\zeta\\
  \end{equation}
  \begin{equation}
    \label{new_def_theta}
    \theta\equiv 1/z\equiv\frac{a'}{a^{d/2}\phi_{0}'},
  \end{equation}
  then sections \ref{scalars} and \ref{grow_decay} immediately
  generalize $d$ dimensions.  Indeed, it is straightforward to show
  that equations (\ref{uv_EOM}) through (\ref{f_dom}) all remain true
  in $d$ dimensions, provided we take $\epsilon$, $u$, $v$, $\theta$,
  and $z$ to be given by their $d$-dimensional definitions
  (\ref{new_def_eps}), (\ref{new_def_uv}) and (\ref{new_def_theta}).

  In particular, this means that the duality extends to
  $d$-dimensions: $u$, $f_{\textrm{grow}}$ and $f_{\textrm{decay}}$
  are all invariant under $\epsilon\rightarrow 1/\epsilon$.  The
  duality pairs expanding solutions in the range
  $\frac{d-4}{d-2}<\epsilon<1$ with contracting solutions in the range
  $1<\epsilon<\infty$.  Note that for $d>4$, there are some
  contracting models with no expanding dual; but for $d=4$, every
  expanding model has a contracting dual, and \emph{vice versa}.

  In the $x\rightarrow 0$ limit in $d$ dimensions, define the growing
  and decaying spectral indices
  \begin{subequations}
    \begin{eqnarray}
      x^{d-1}|f_{\textrm{grow}}|^{2}
    & \propto & x^{n_{\textrm{grow}}-1}\\
      x^{d-1}|f_{\textrm{decay}}|^{2}
    & \propto & x^{n_{\textrm{decay}}-1}.
    \end{eqnarray}
  \end{subequations}
  Then, using (\ref{def_alpha}), (\ref{hankel_small_x}),
  (\ref{f_sub}), (\ref{def_gamma}) and (\ref{f_dom}) we find
  \begin{subequations}
    \begin{eqnarray}
      n_{\textrm{grow}}\;-1 & = & 
      d-3-\;\Big|\frac{\epsilon+1}{\epsilon-1}\Big|\\
      n_{\textrm{decay}} -1 & = & 
      d-1-\left|\Big|\frac{\epsilon+1}{\epsilon-1}\Big|-2\right|
    \end{eqnarray}
  \end{subequations}
  with $\epsilon$ given by (\ref{new_def_eps}).  Note that these
  expressions are invariant under $\epsilon\rightarrow 1/\epsilon$.

  As shown in \cite{Kodama_Sasaki}, we may again introduce tensor
  perturbations through eq. (\ref{tensor_perturbed_metric}).  Now
  define the gauge-invariant variable $f_{T}^{(d)}\equiv
  a^{(d-2)/2}h_{T}$ which obeys
  \begin{equation}
    \label{new_f_EOM}
    f_{T}''+[k^{2}-(a^{(d-2)/2})''/a^{(d-2)/2}]f_{T}=0
  \end{equation}
  along with the boundary condition (\ref{tensor_bc}).  Using
  (\ref{tensor_bc}), (\ref{new_background_soln}) and (\ref{new_f_EOM})
  we find that solution for $f_{T}$ is still given by (\ref{f_soln})
  in $d$ dimensions.  In the $k\rightarrow 0$ limit, define the tensor
  spectral index $k^{d-1}|f|^{2}\propto k^{n_{T}}$.  Then, using
  (\ref{hankel_small_x}) and (\ref{f_soln}) we find
  \begin{equation}
    n_{T}=d-1-\left|\frac{\epsilon-3}{\epsilon-1}\right|
  \end{equation}
  with $\epsilon$ given by (\ref{new_def_eps}).  Note that this
  expression is \emph{not} invariant under $\epsilon\rightarrow
  1/\epsilon$.  This completes the generalization to $d$ spacetime
  dimensions.

\end{document}